\documentclass[smallextended]{svjour3}       
\smartqed  
\usepackage{graphicx}
\usepackage{hyperref}
\usepackage{gensymb,siunitx} 
\usepackage{booktabs}
\usepackage{cite}
\usepackage{lipsum}
\usepackage{mathtools}
\usepackage{geometry}
\usepackage{ braket,  array}

\begin{document}

\title{Analysis of Single Photon Detectors in Differential
Phase Shift Quantum Key Distribution
}


\author{Vishal Sharma         \and
}


\institute{Vishal Sharma \at
bits.vishal11@gmail.com \\
\newline
\newline
Instrumentation and Applied Physics \\
Indian Institute of Science, Bengaluru \\
CV Raman Road, Bengaluru, Karnataka-560012 \\
           \and
           \at
          \\
}

\date{Received: date / Accepted: date}

\maketitle

\begin{abstract}
	In the current research work, an analysis of differential phase shift quantum key distribution using InGaAs/InP and Silicon-APD (avalanche photodiode) as single photon detectors is performed. Various performance parameters of interest such as shifted key rate, secure key rate, and secure communication distance obtained are investigated. In this optical fiber-based differential phase shift quantum key distribution, it is observed that Si-APD under frequency conversion method at telecommunication window outperforms the InGaAs/InP APD.
\end{abstract}

\section{Introduction}
\label{intro}
Quantum key distribution (QKD) shares the secure secret key among the authenticated users, where unconditional security is achieved by the postulates of quantum mechanics, and different from the classical cryptography where computational complexity is the basis for the entire cryptography system. Many research communities \cite{lutkenhaus2000security, waks2002security, sharma2022analysis} performed  security tests and detailed analysis under the realistic scenarios, and concluded that source characteristics such as single or entangled photons are one of the performance deciding factors for any quantum-cryptography system. 
Quantum key distribution was first implemented in 1992 \cite{bennett1992experimental} and desired improvements were developed in \cite{sharma2019analysis,sharma2020quantum, sharma2018analysis, gobby2004quantum, sharma2022analysis, sharma2015controlled, sharma2016comparative, sharma2018decoherence, sharma2018quantum, raj2019quantum}.Quantum technologies are nowadays being deployed in many interdisciplinary industrial applications \cite{sharma2021quantum, sharma2014feasibility,sharma2015experimental}. Wavelength at 1550\,\, nm is the desired one for practical deployment of quantum communication, as it provides less losses (0.2 dB/km) as compared to 1300\,\, nm wavelength which offers higher losses (0.35 dB/km). There are various single photon based quantum key distribution protocols implemented experimentally such as Bennett-Brassard 1984 (BB84) protocol, the entanglement-based Bennet-Brassard-Mermin 1992 (BBM92) protocol \cite{honjo2004differential}. In the present work, we consider differential-phase-shift quantum key distribution (DPS-QKD) protocol \cite{inoue2003differential,sharma2016effect},  deploying weak coherent pulse train \cite{inoue2002differential, inoue2003differential}, implemented under  optical-fiber-based experimental parameters based on InGaAs/InP and silicon-APD at  telecommunication wavelengths. For efficient detection of single-photons at 1550\,nm, we use frequency-up conversion technique \cite{langrock2005highly}. We use silicon-APD due to its unique properties and advantages such as high quantum efficiency, low dark counts rates with high timing accuracy and excellent timing stability, with suitable wavelength conversion to 1550\,nm \cite{IDQ,pelc2012cascaded,sharma2019analysis} and tuning the experimental parameters to provide very low losses and provides higher secure key rate (SKR) \cite{sharma2014analysis} as compared to InGaAs/InP-APD \cite{aureatechnology}. In addition to these, further, we analyze vulnerability of DPS-QKD under different hybrid attacks.

\section{Single Photon Detectors at Third Telecommunication Window}

\subsection{Single photon detectors}
Single-photon detection in optical fiber-based quantum key distribution (QKD) has been investigated in many applications where InGaAs/ InP avalanche photodiodes were used due to their experimental properties in QKD systems  \cite{yoshizawa200410, bethune2004high, stucki2001photon, bourennane2001single, gobby2004unconditionally}. Further, it was analyzed that due to trapped charge carriers, these detectors suffer from after-pulse effects, low quantum efficiencies and hence results in relatively large dark count rates. These drawbacks degrade the performance of InGaAs/InP avalanche photodiodes in gated-mode operation. Under such conditions (in gated mode), the detector works above the breakdown threshold for a limited duration which indicates performance improvement in terms of high photon detection efficiency with comparatively less dark counts.
After a short interval of time, it returns below breakdown for the  time duration enough for the trapped charge carrier to leak away. In gated mode operation, the device is allowed to work at mega-hertz rates, for which trapping lifetime spans microsecond timing. Hence, we are able to reduce the after pulse probability by the fraction of gate width to the time separation between gates. At this point, it is very important to notice the significance of the gate frequency which is one of the performance deciding factor in almost all type of QKD applications, which decides the pulse repetition rate, and further limits the achievable communication rate. In addition to this, semiconductor material's response time is one of the important parameters which decides the rate of dark counts produced and responsible for limiting the communication distance achieved, and moreover, all these are affected by gate width. In general,  gate widths of 1-2\,\, ns at  $\sim$ 1 MHz pulse repetition frequency are deployed with final dark counts  of $10^{-5}$ /gate order. In the current analysis of DPS-QKD, we are deploying two single photon detectors, Si-APD \cite{IDQ} and InGaAs/ InP APDs.

\subsection{Single-Photon Detection with Frequency Up-conversion}
A periodically poled lithium niobate (PPLN) is deployed for sum-frequency generation \cite{roussev2004periodically}, and a strong pump at $1320\,\, nm$ is allowed to interact with a single photon at $1550\,\, nm$. This technique is applied in the $1550\,\, nm$  up-conversion single-photon detector \cite{langrock2005highly}. With the use of  the PPLN waveguide, it becomes possible to convert a signal of  very high conversion efficiency  to  a  $700\,\,nm$ sum-frequency output. This is achieved due to the  guided wave structure  of PPLN waveguide, where the presence of  quasi-phase-matching pattern and the tight mode confinement over longer  interaction lengths exist.
Further, by applying these methods, silicon APD  detects the converted  information carriers (photons). We are deploying two single photon detectors, Si-APD, and  InGaAs/ InP APDs. Out of these two detectors, Si-APD is mostly preferred in all industrial and real-field QKD (Quantum Key Distribution ) applications. This is because of its high quantum efficiency in the NIR (near infrared region), low after-pulse effects, and low dark-count rates, low dead time(45\,ns), and  a timing resolution as low as (40 \,ps) \cite{IDQ}. These  unique characteristics of  silicon APD outperforms  InGaAs/ InP APDs  in almost all QKD applications where single-photon detection is achieved with increased value of timing accuracy and stability up to a count rate of 20 MHz.  \cite{IDQ, sharma2019analysis, pelc2012cascaded}. The Geiger mode characteristic of Si-APD, which is also known as nongated mode of operation, is based on low-after-pulse probability, and helps in achieving higher communication rate in practical QKD systems. Further, dead time  of Si-APD diminishes the secure key generation rate, here in Si-APD, it is  low dead time(45\,\, ns). During this time period that further gives a photodetection event, the photodiode cannot reply  to next occurrences, and, finally, a  larger amount of  photon flux saturates the set-up.
In the up-conversion process, both the quantum efficiency $\eta_{up}$ and the dark-count rate $D_{up}$ depend on pump power $p$ \cite{langrock2005highly}. In Si-APD, $0.46$ value of quantum efficiency is obtained, when the $100\%$  photon  conversion condition is met, it is possible in the case when the phase-matching condition in the waveguide is met and enough  pump power is present  for the said process. In a waveguide, due to three-wave interactions based on coupled-mode theory, the fitting curve is obtained from the following relation 

\begin{equation}
\eta_{up}(p)  = a_{1} \sin^{2} (\sqrt {a_{2} p}),
\end{equation}

where $p$ is in mW, and values of $a_{1}$, and $a_{2}$ are 0.465, and 79.75, respectively.

We believe that the dark-count rate is controlled  by the below mentioned  nonlinear process. At first, the pump photons are  dispersed  by the   fiber  and phonons of  the PPLN waveguide via a spontaneous Raman scattering mechanism. This method escalates straightaway with the pump power, and produces a spread of Stokes photons, which contains 1550\, nm  signal wavelength . Afterwards, the noise photons combine with the pump photons in the waveguide via the phase-matched sum-frequency generation approach, and generate dark counts. The combined process produces a precise  quadratic dependence of the dark counts on the pump power, as shown in Figure $ \ref{fig:PumpPowerVSDarkCount_21July}$. The following expression generates an accurate polynomial fitting curve 
\begin{equation}
D_{up}(p) = b_{0} + b_{1} p + b_{2} p^{2} + b_{3} p^{3} + b_{4} p^{4}, \,\,\,\,\,\, (s^{-1}),
\end{equation}

where the values of $b_{0}$, $b_{1}$, $b_{2}$, $b_{3}$, and $b_{4}$ are 50, 826.4, 110.3, -0.403, and 0.00065, respectively. The value of power $p$ is  given in $mW$.

Dark counts are also generated by the parametric fluorescence process and up-conversion of noise signal photons  [39, 40]. Here, we find the quadratic relation between the dark counts and associated pump power. In such fluorescence processes in the frequency conversion detector, 8.9-$\mu$ m idler photons are absorbed in lithium niobate, and hence more dark counts are produced because of the  spontaneous Raman scattering and the the process described. Further, these undesired dark counts can be minimized by interchanging the signal wavelengths and pump [12], which is the important process generated in a waveguide, where thermal process of excited vibrational states produces anti-Stokes scattering gain.

Dark counts limit the performance of the up-conversion detector. The waveguide bandwidth decides the number of dark counts which is also responsible for the number of noise photons. For a detector with $B_{d}$ bandwidth, the parameter $D_{up}\,\,_{Hz}$ is defined as  $D_{up}\,\,_{Hz} = D_{up}/B_{d}\,\,\,s^{-1} Hz^{-1}$, where the term $B_{d}$ is the dark count per mode. In general, an ideal communication system can be taken into consideration in which the  bit rate $B$ is equal to the bandwidth $B$ with a matched filter. Here the measurement time window for such an ideal communication system is $1/B$. Based on similar concepts, performance of quantum key distribution systems are based on a very important parameter known as dark counts per time window, $d_{up}$ is equal to $D_{up}$ Hz. Further, considering the optimum filtering case, $d_{up}$ is independent of the bit rate $B$. In InGaAs/ InP APD  in gated mode operation, with $1/B$ gate width, and $d_{APD}$  is the dark counts per gate, which is computed by  $D_{APD}/B $, the term $D_{APD} (s^{-1})$ is known as the dark-count rate of the InGaAs/ InP APD. In InGaAs/ InP APD, $D_{APD} = 10^{4}, \,\,\,\, s^{-1}$ is used. Table I represents  the dark-count quantities used in the current research work, where bit rate is denoted by symbol $B$ and the waveguide bandwidth is represented by $B_{d}$. In the up-conversion process, the normalized noise equivalent power(NEP)  $\frac{ \sqrt{2D_{up}} }{\eta_{up}}$ is reduced, which is related to $D_{up} = 6.4 \times 10^{3} \,\,\,\, s^{-1}$ and $\eta_{up}= 0.075 $. Here the parameter $D_{up}$ Hz is computed at the operating point of the detector. For one of the cases, if the bandwidth,  $B_{d} = 50$ GHz  for an up-converter detector, which results in $1.3 \times 10^{-7}$ as an optimum value of $d_{up}$. These manipulations are prior estimations that play an important role in many practical QKD applications for deciding QKD performance parameters of interest. 
Frequency-up conversion methods used in quantum communication systems improve the detection performance at telecommunication  wavelength. Further, waveguide bandwidth also affects the number of dark counts in Si-APD and its characteristics also depend on pump power. We investigate all these factors in detail in the coming sections. 

\section{Differential Phase Shift Quantum Key Distribution protocol}
DPS-QKD possesses many non-orthogonal states with many pulses, as shown in Fig. 1 \cite{inoue2003differential}. These pulses, in highly attenuated coherent states are randomly phase modulated $\{0, \pi\}$. Bob applies random modulation on the delay time, $NT$, where $N$ is a positive integer, and $T$ is the reciprocal of the clock frequency, detector clicks based on the phase difference of the two pulses which are having a $NT$ time difference.Bob announces the value of $N$  and the time instances on which the photon was detected. Alice comes to know which detector clicked.Based on these events, Alice and Bob assign the bit values to the detectors. This protocol is protected from  the individual attacks because the two non-local pulses are used for information encoding in terms of the difference of their respective phase information.

\begin{figure}[!t]
\centering
\includegraphics[width=6.0in]{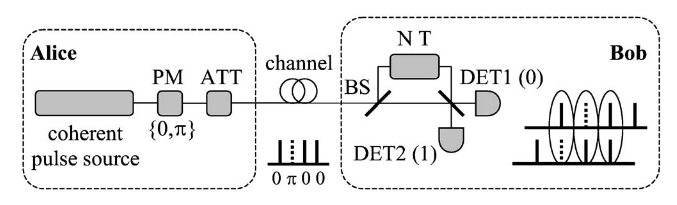}
\caption{DPS-QKD protocol \cite{inoue2003differential}. Phase Modulator (PM), Attenuator(ATT), Beam Splitter(BS) and Detector(DET).}
\label{fig_sim}
\end{figure}

Based on the number of detected photons, we calculate the sifted key rate, and the secure rate is manipulated from the photon splitting and general individual attacks.

Further, we perform the security analysis of DPS-QKD under some of the well-known eavesdropping attacks such as Beam-splitter attack  and Intercept-resend attack. Due to the fact that, the encoded information of the differential phase of two successive nonlocal pulses is transmitted which provides the secure protocol and the protocol becomes robust against the said attacks \cite{inoue2005robustness, honjo2007long}. To investigate the secure communication rate under the said attacks we need to find out the privacy amplification shrinking factor ($ \tau $) with respect to average collision probability $(p_{c})$  under various attacks.

\begin{table}[h!]
\centering
\caption{Dark count parameters for frequency up-conversion.}
\scalebox{1.50}{
\begin{tabular}{|c c c|} 
 \hline
  & $InGaAs/InP$\, APD & Up-converter\\ 
 \hline
 Dark count rate ($s^{-1}$)  & $D_{APD}$ &   $D_{up}$ \\ [0.1 ex]
 Dark counts per mode  $(s^{-1} Hz ^{-1})$  & - &  $D_{up} Hz = \frac{D_{up}}{B_{d}}$ \\
 Dark counts per time window/gate &  $d_{APD} = D_{APD} \frac{1}{B}$ &  $d_{up} = D_{up} Hz$\\ [0.1 ex]
 \hline
\end{tabular}}
\end{table}


\begin{table}[h!]
\begin{center}
\centering
\caption{Error-correction algorithm given in \cite{brassard1994advances}}
\label{tab:table1}
\scalebox{3.00}{
\begin{tabular}{|c  c|} 
\hline
$e$  &   $f(e)$  \\ [0.5ex] 
\hline
0.01  &   1.16\\
0.05  &   1.16 \\
0.1  &   1.22 \\
0.15   &  1.35 \\  [1.0ex] 
\hline
\end{tabular}}
\end{center}
\end{table}

Bob's photon detection probability is expressed as 
\begin{equation}
p_{click} = p_{signal} + p_{dark},
\end{equation}
where, 
\begin{equation}
p_{signal} = \mu \eta 10^{-(\alpha L + L_{r})/10},
\end{equation}

\begin{equation}
p_{dark} = 2d,
\end{equation}

where $\mu$ is the mean photon number per pulse, $\alpha$ is the optical fiber loss coefficient in dB/km, $\eta$ denotes the detector quantum efficiency, $L_{r}$ is the losses in the receiver unit, $L$ is the communication link in km, between the two authentic users, say Alice and Bob, $d$ denotes the dark counts per measurement time window. In $p_{dark}$, $2$ denotes the number of detectors used at the Bob's detection unit. In the ideal case $\mu = 1$, and in a Poisson source, $\mu$ is a free variable which has to tune for optimum performance \cite{gisin2002quantum}.

\begin{equation}
e =  \frac{(\frac{1}{2} p_{dark}+ b\,\, p_{signal})}{p_{click}},
\end{equation}
where, $e$ and $b$ denote the error rate and baseline system error rates, respectively.

The value of $f(e)$ is based on the error-correction algorithm as mentioned in Table II \cite{brassard1994advances}.

In privacy amplification procedure, the main shrinking factor  $\tau(e, \beta)$ is written as 

\begin{equation}
\tau = -log_{2}p_{c}
\end{equation}
where $p_{c}$ is the average collision probability, which shows the amount of  Eve's mutual information with Bob and Alice.
 
The following expression is generated for $\tau$.

\begin{equation}
\tau(e, \beta) = - \beta log_{2} \Big[\frac{1}{2} + 2\Big( \frac{e}{\beta}\Big) -2\Big(\frac{e}{\beta}\Big)^{2} \Big]
\end{equation}

Source emits photons where fraction of single-photon states is written as 

\begin{equation}
\beta = \frac{p_{click} - p_{m} } {p_{click}},
\end{equation}

where the $p_{m}$ is associated with the probability of the multi-photon quantum states. In case of an ideal single-photon state, $p_{m}$ = 0 or $\beta = 1$. In another type of source the value of $p_{m}$ , e.g. the probability of photon emission in  Poisson source is expressed as 

\begin{equation}
p_{m} = 1 - (1+\mu) e^{-\mu}
\end{equation}

The term $\beta$ decides the photon number splitting (PNS) attack. Eve makes PNS attack to steal the information fully or partially decided by the term $\beta$. Her  quantum nondemolition measurements of the photon number in each pulse hide her presence in between the communication line without producing any significant errors. As the source emits multiple photons, Eve takes advantage of that by taking one photon in quantum memory, and performing a delayed quantum measurement on the photon after the basis announcement made by Bob. The PNS attack is the main obstacle that limits the performance of the laser source used in BB84 QKD protocol experiments. The loss in the secure communication rate with the communication distance, $10^{\frac{- \alpha L}{10}}$, for small error rate and $p_{dark} \ll p_{signal} \ll 1 $. In another case, under the same operating conditions, when the deployed source is an ideal single-photon source, it is observed that  $R_{BB84} \approx  \frac{1}{2} \nu p_{signal}$. 

Here in all the security analysis we assume that Eve is technically sound and possesses a quantum memory with infinitely long coherence time because the authentic users (here Alice and Bob) can announce the basis measurement related outcomes with long delay. On the other hand, if Eve does not have such a quantum memory she has to perform polarization measurement with arbitrary random basis selection. In such case, we can write Equation (8) as follows

\begin{equation}
\tau(e, \beta) = - \frac{1+\beta}{2} log_{2} \Big[ \frac{1}{2} + 4 \Big(\frac{e}{1+\beta}\Big) - 8\Big(\frac{e}{1+\beta}\Big)^{2}  \Big]
\end{equation}

BB84 protocol can be designed to be robust against photon number splitting attack by deploying decoy states \cite{lo2005decoy, ma2005practical, wang2005beating}, or altering the sifting procedure \cite{scarani2004quantum}. In addition to these, secure communication distance is achieved by using Poisson sources in BB84 QKD protocol.

\begin{figure}[!t]
\centering
\includegraphics[width=6.000in]{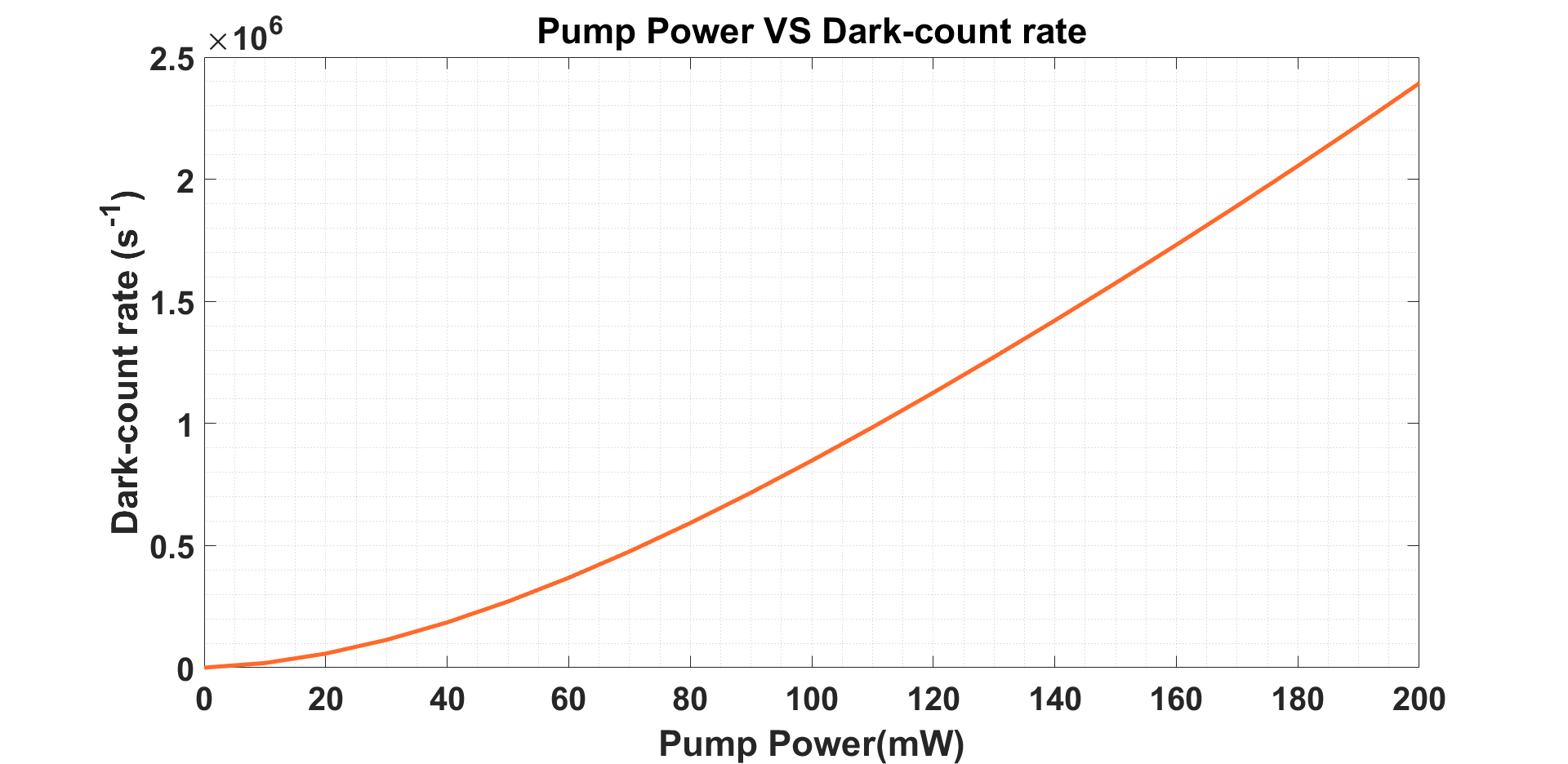}
 \caption{Dark count dependency on the applied pump power.}
\label{fig:PumpPowerVSDarkCount_21July}
\end{figure}

\begin{figure}[!t]
\centering
\includegraphics[width=6.000in]{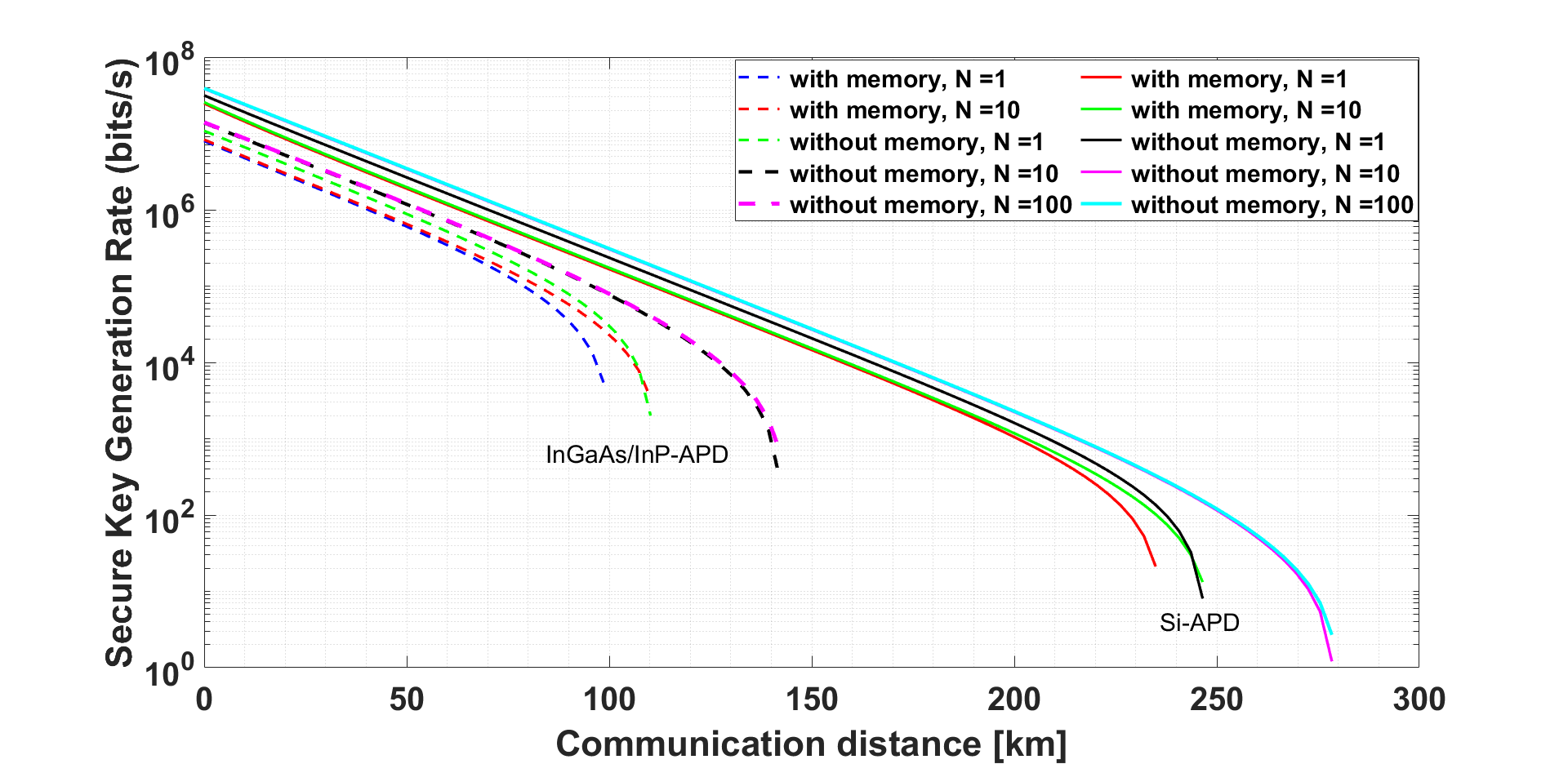}
\caption{\label{Rdpsqkd_24JUNE2022.png} Secure key generation under the considered attacks, $b=0.01$; $\mu=0.2$; $f=1.16$; $\nu=1*10^{9}$; $\eta_{1}=0.155$; $\eta_{2}=0.35$; $\alpha=0.21$; $L_{r1}=3.0$; $L_{r2}=2.1$; $d_{1}=9.2*10^{-6}$; $d_{2}=3.5*10^{-8}$; $N_{1}=1$; $N_{2}=10$; $N_{3}=100$; $t_{d1}=200*10^{-9}$; $t_{d2}=45*10^{-9}$. Here subscript 1 and 2 denote InGaAs-APD and Si-APD, respectively.}
\end{figure}

\begin{figure}[!t]
\centering
\includegraphics[width=6.000in]{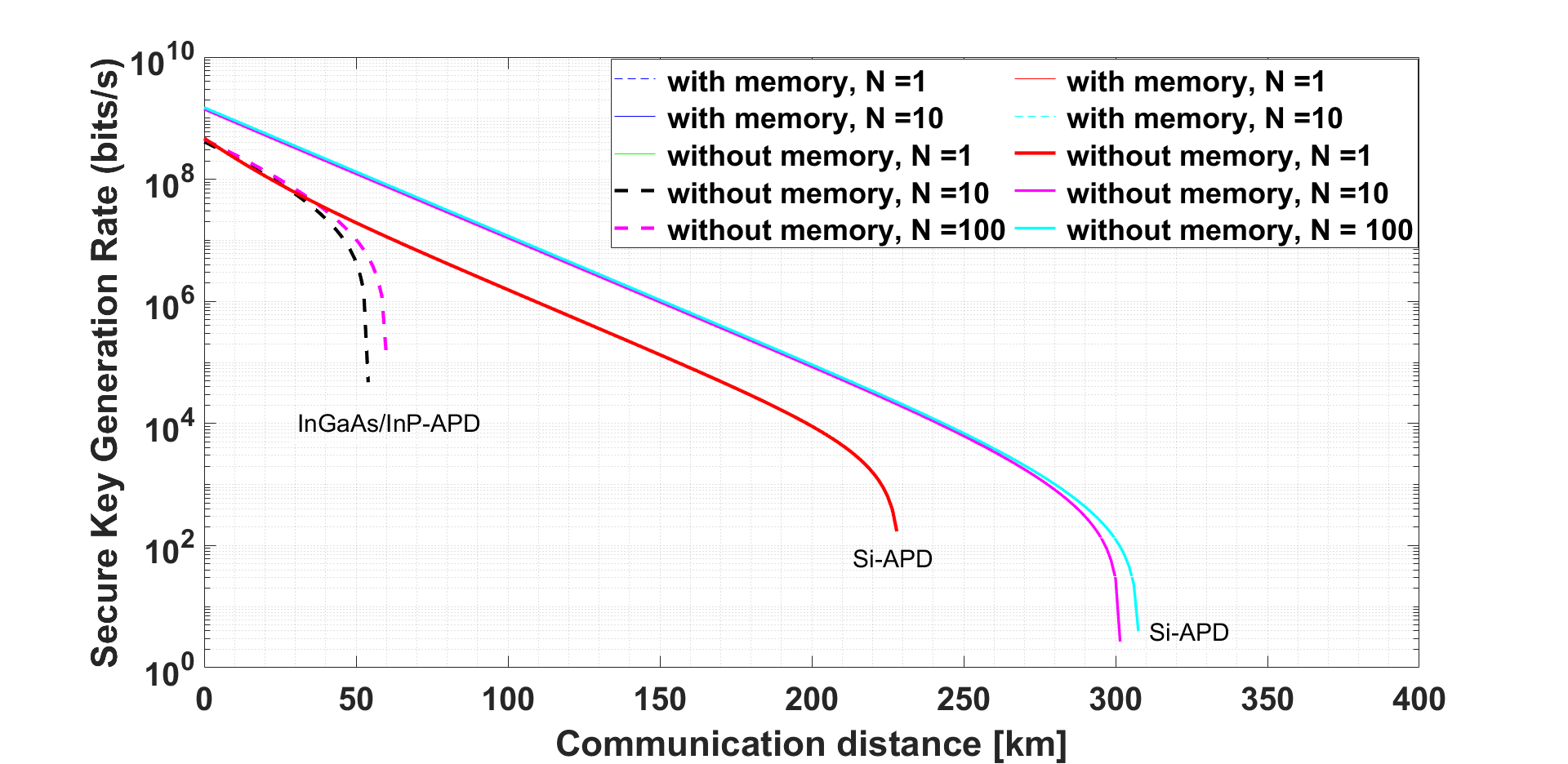}
\caption{\label{Rdpsqkd_24JUNE2022.png} Secure key generation under the considered attacks, $b = 0.01$; $\mu = 0.77$; $f=1.16$; $\nu=10*10^{9}$; $\eta_{1}=0.155$; $\eta_{2}=0.35$; $\alpha=0.21$; $L_{r1}=3.0$; $L_{r2}=2.1$;  $d_{1}=2.0*10^{-3}$; $d_{2}=3.5*10^{-8}$; $N_{1}=1$; $N_{2}=10$; $N_{3}=100$; $t_{d1}=200*10^{-9}$; $t_{d2}=45*10^{-9}$. Here subscript 1 and 2 denote InGaAs-APD and Si-APD, respectively.}
\end{figure}

\subsection{Beam-splitter attack}

\begin{figure}[!t]
\centering
\includegraphics[width=6.000in]{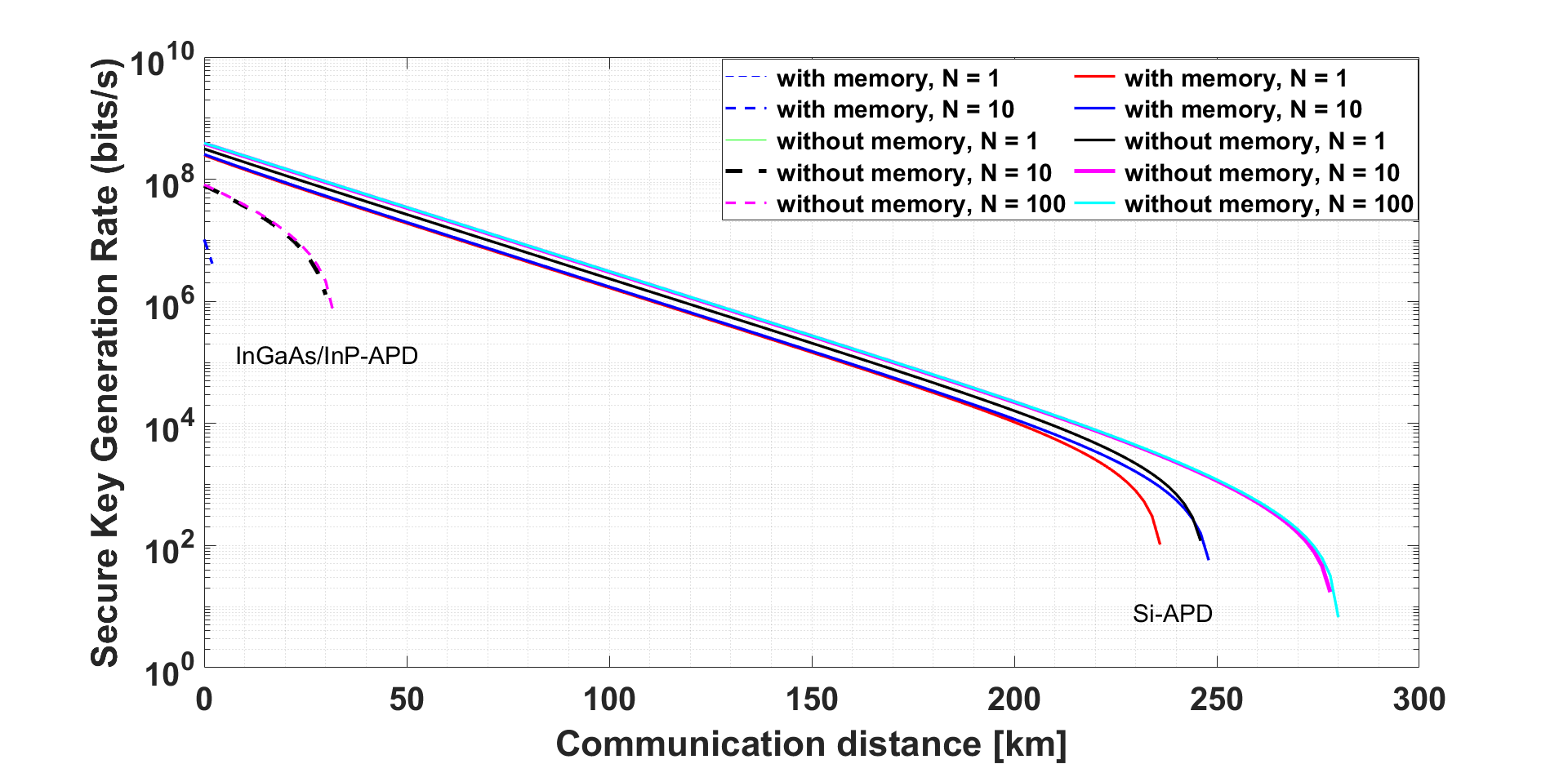}
\caption{\label{Rdpsqkd_24JUNE2022.png} Secure key generation under the considered attacks, $b = 0.01$; $\mu = 0.2$; $f=1.16$; $\nu=10*10^{9}$; $\eta_{1}=0.155$; $\eta_{2}=0.35$; $\alpha=0.21$; $L_{r1}=3.0$; $L_{r2}=2.1$;  $d_{1}=2.0*10^{-3}$; $d_{2}=3.5*10^{-8}$; $N_{1}=1$; $N_{2}=10$; $N_{3}=100$; $t_{d1}=200*10^{-9}$; $t_{d2}=45*10^{-9}$. Here subscript 1 and 2 denote InGaAs-APD and Si-APD, respectively.}
\end{figure}

Alice transmits multiple photons to Bob, these photons are intercepted by Eavesdropper to get the replica of the coherent quantum states. In this strategy, Eve deploys a beam splitter with transmission $\eta_{BS}$. There are other possibilities, where Eve uses lossless fiber in place of lossy fiber. In addition to these,  Eve replaces inefficient detectors by ideal detectors at the receiver end (at Bob's end). The probability, $p_{signal}$, is equivalent to Eq. (5), and is known as Bob's signal photon detection probability. Eve tries to unchange this probability value so that she can hide her presence, for that she changes the value of the beam-splitter transmission  $\eta_{BS}$  to

\begin{equation}
 \eta_{BS} = \eta \,\, 10^{-(\alpha L + L_{r})/10},  
\end{equation}

The necessary parameters for the above equation is already mentioned. Eve at this stage can use an interferometer with $M \tau$ delay time selected randomly from Bob's, to get insight from the intercepted pulses. The amount of information measured can be calculated as follows. The expressions for the detection probability at Bob's and Eve's end for a given time slot are written as $ \mu(1- \eta_{BS})$ and $\mu \eta_{BS}$, respectively. The term $\mu$ represents the value of mean photon number. Further, the detection probability at the same time instance is expressed as $ \mu^{2} \eta_{BS} (1- \eta_{BS})$. Analyzing further, it is observed from the concept of conditional probability that the value of probability for an Eavesdropper received a particular bit at a given point of time where Bob has already detected the photon at that particular time frame is written as $ \mu^{2} \eta_{BS} (1- \eta_{BS})/\mu \eta_{BS} = \mu (1- \eta_{BS}) $. The expression $\frac{1}{N}$ represents the probability value that the Eavesdropper randomly selected $M$ overlaps  Bob's $N$. At this stage, it can be written that the probability value achieved by Eavesdropper in comparison to Bob is $\mu(1 - \eta_{BS})/N$. This expression for the obtained probability by Eavesdropper is valid only and only when she is not well equipped with a quantum memory for infinitely long coherence time. On the other hand, Eavesdropper can be equipped with a quantum memory to store the incoming photon pulses and by the time she listens to Bob's announcement to frame her strategy accordingly. To be successful in her information-gaining strategy, Eve should possess a quantum memory with long coherence time, as the authentic users, Alice and Bob, can randomly delay their individual announcements. Here, Eavesdropper deploys an optical switch with a suitable interferometer in place of a beam splitter to access the pulses for which Bob already gained the differential phase information . Hence, using this technique, there is a significant information gain to Eve, which is equal to $2\mu(1 - \eta_{BS})$. Finally, using this attack strategy, Eavesdropper obtains $p_{c} = 1$ amount of information which corresponds to the fraction of bits equal to $\mu(1- \eta_{BS})/N$ or $2 \mu (1 - \eta_{BS})$. Here the BS (Beam-Splitter) attack does not introduce errors in the two authentic communicating parties, Alice and Bob. The rest of the bit fractions are given by 
(i) In \,\,  the \,\,  absence \,\, of \,\, quantum \,\, memory
\begin{equation}
\gamma_{1} = 1 - \mu \frac{(1 - \eta_{BS})}{N} = 1- \frac{\mu}{N} +\frac{p_{signal}}{N},
\end{equation}
(ii) In \,\, the \,\, presence \,\, of  \,\, quantum \,\, memory
\begin{equation}
\gamma_{2} = 1 - 2\mu (1 - \eta_{BS}) = 1- 2\mu +2p_{signal}, 
\end{equation}

\begin{figure}[!t]
\centering
\includegraphics[width=6.000in]{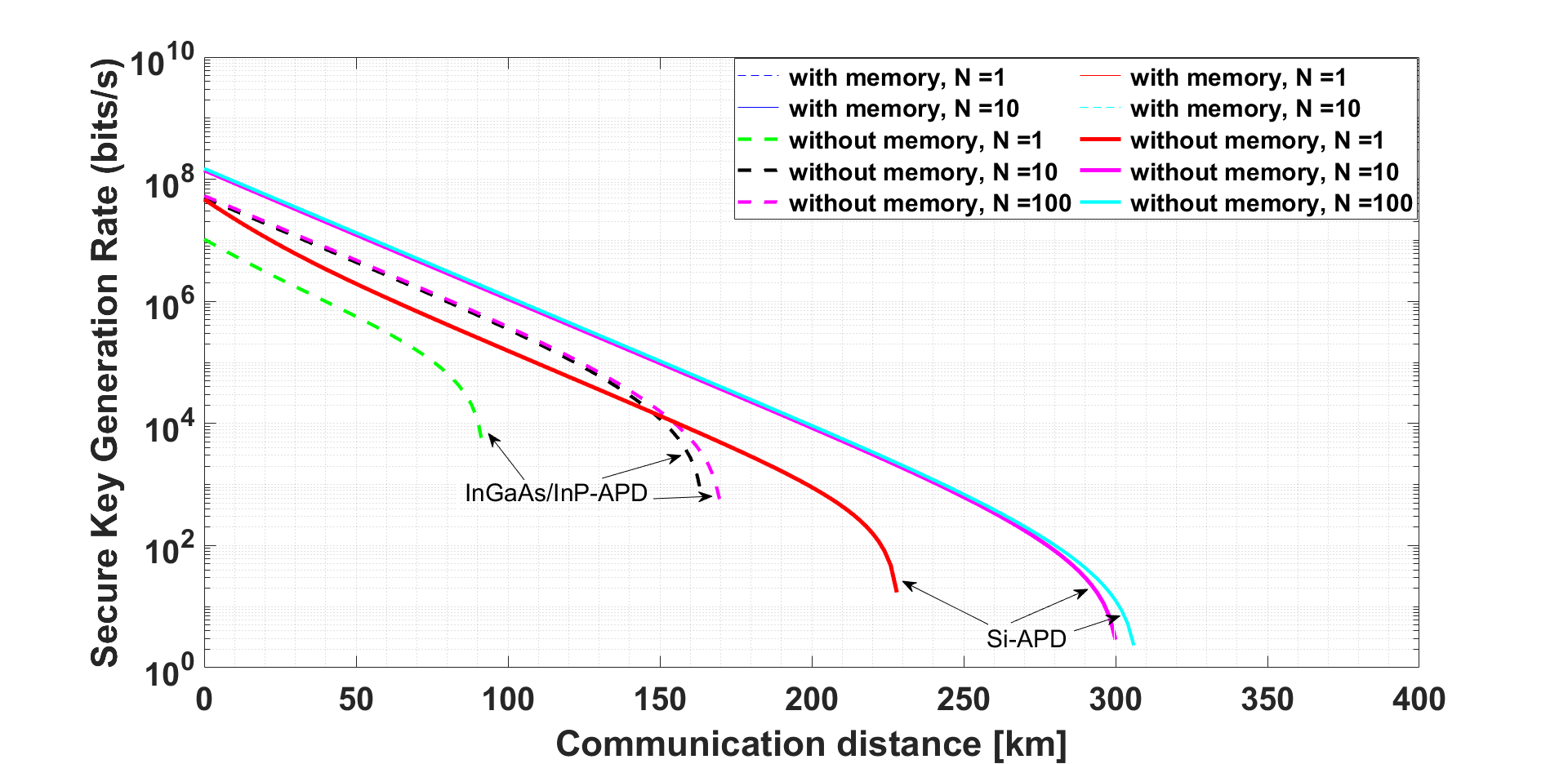}
\caption{\label{Rdpsqkd_24JUNE2022.png} Secure key generation under the considered attacks, $b = 0.01$; $\mu = 0.77$; $f=1.16$; $\nu=1*10^{9}$; $\eta_{1}=0.155$; $\eta_{2}=0.35$; $\alpha=0.21$; $L_{r1}=3.0$; $L_{r2}=2.1$; $d_{1}=9.2*10^{-6}$; $d_{2}=3.5*10^{-8}$; $N_{1}=1$; $N_{2}=10$; $N_{3}=100$; $t_{d1}=200*10^{-9}$; $t_{d2}=45*10^{-9}$. Here subscript 1 and 2 denote InGaAs-APD and Si-APD, respectively.}
\end{figure}

\subsection{Intercept-resend attack}

Eavesdropper applies Intercept-resend attack which is  another strategy applied on the photon pulses being sent from Alice to Bob. In such a strategy, $MT$ time difference based two pulses are hacked by the Eavesdropper, they are passed via an interferometer with the same delay, further differential phase measurement takes place, and based on the outcomes Eve sends it to Bob. By this strategy, Eve splits a single photon into two with correct phase difference which is difficult to detect her presence, and hence Bob considers it as sent correctly by Alice. In this way, the Eavesdropper hides her presence. Bob can detect her presence when he measures with the probability value $(1 - 1 / 2N)$ and  the delay value, $N \neq M$. Hence, this gives the error value equivalent to $1/2(1 - 1 / 2N)$. In such a situation, Eve can try  the value $2e /(1 - 1/2N)$ of the pulse pair, which is less than the error rate, where $e$ is the error rate. For obtaining full information by the Eavesdropper, the probability value is $1/2N$.

\begin{figure}[!t]
\centering
\includegraphics[width=6.000in]{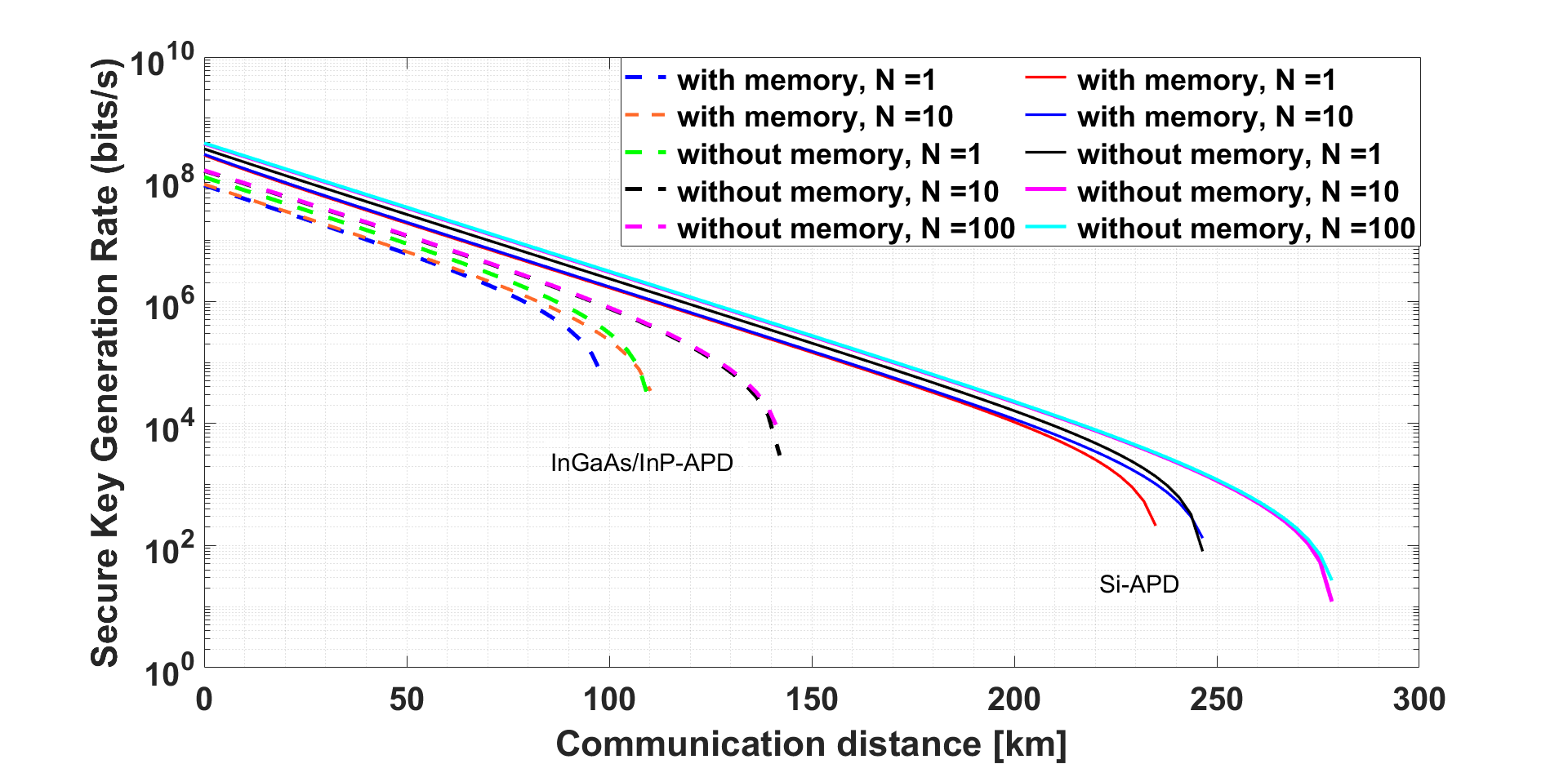}
\caption{\label{Rdpsqkd_24JUNE2022.png} Secure key generation under the considered attacks, $b = 0.01$; $\mu = 0.2$; $f=1.16$; $\nu=10*10^{9}$; $\eta_{1}=0.155$; $\eta_{2}=0.35$; $\alpha=0.21$; $L_{r1}=3.0$; $L_{r2}=2.1$; $d_{1}=9.2*10^{-6}$; $d_{2}=3.5*10^{-8}$; $N_{1}=1$;$N_{2}=10$; $N_{3}=100$; $t_{d1}=200*10^{-9}$; $t_{d2}=45*10^{-9}$. Here subscript 1 and 2 denote InGaAs-APD and Si-APD, respectively.}
\end{figure}

In case of the combined attacks (beam-splitter and intercept-resend attacks), Eve is unknown to the bits $p_{c}= \frac{1}{2}$ which are equal to $\gamma - \frac{e}{N(1-1/2N)}$. In this case, privacy amplification shrinking factor is computed as follows
\begin{equation}
\tau(e, \gamma) = \gamma - \frac{e}{N(1-1/2N)}
\end{equation}

\begin{figure}[!t]
\centering
\includegraphics[width=6.000in]{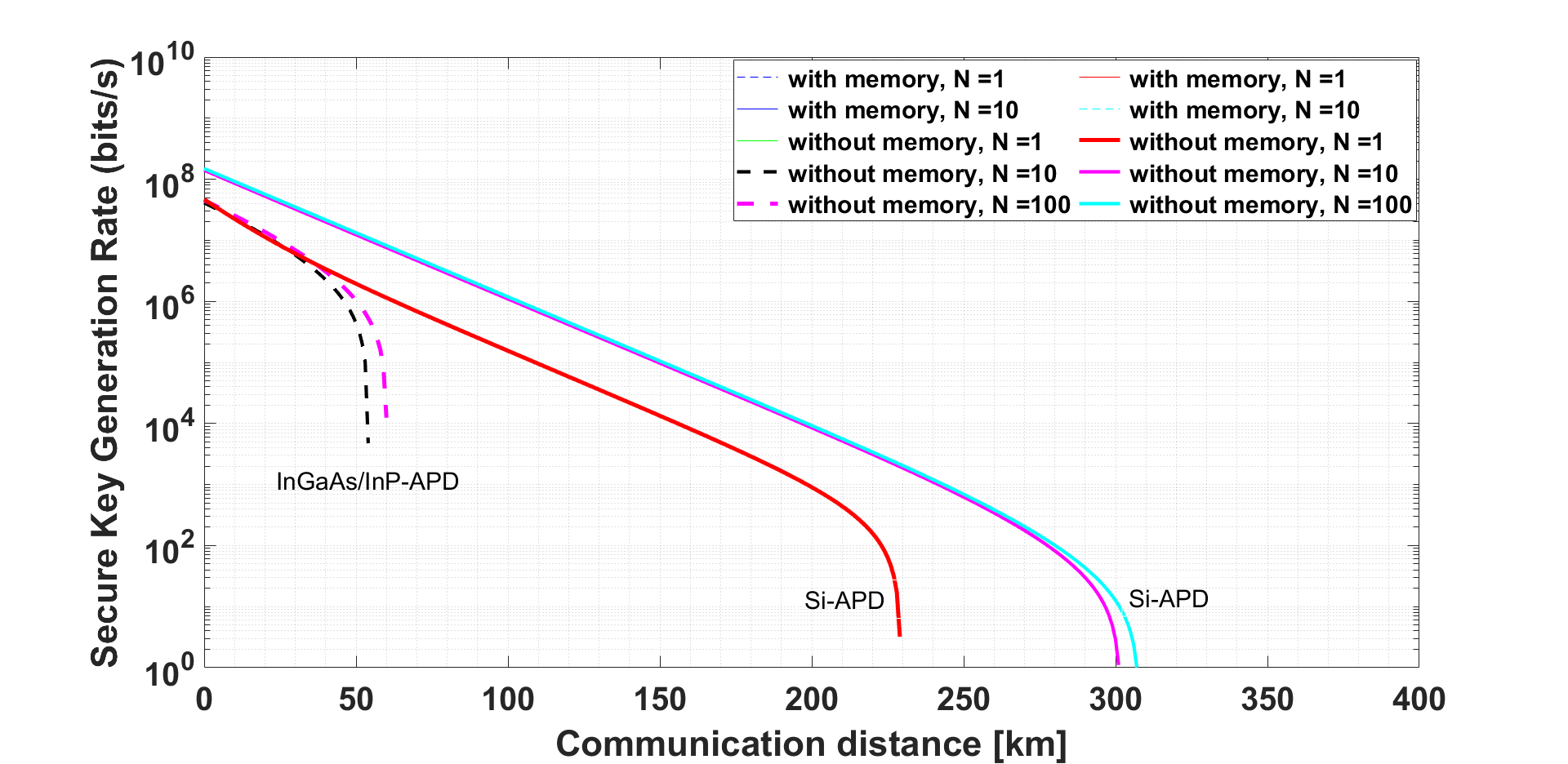}
\caption{\label{Rdpsqkd_24JUNE2022.png} Secure key generation under the considered attacks, $b = 0.01$; $\mu = 0.77$; $f=1.16$; $\nu=1*10^{9}$; $\eta_{1}=0.155$; $\eta_{2}=0.35$; $\alpha=0.21$; $L_{r1}=3.0$; $L_{r2}=2.1$;  $d_{1}=2.0*10^{-3}$; $d_{2}=3.5*10^{-8}$; $N_{1}=1$; $N_{2}=10$; $N_{3}=100$; $t_{d1}=200*10^{-9}$; $t_{d2}=45*10^{-9}$. Here subscript 1 and 2 denote InGaAs-APD and Si-APD, respectively.}
\end{figure}

Here Equations 13 and 14 are used to calculate $\gamma$. Now we can express the secure communication rate equation for Differential Phase-QKD under the combined attack (beam-splitter and intercept-resend attacks).

\begin{equation}
R_{dpsqkd} = \nu \, p_{click} \{\tau(e, \gamma) +f(e) [e log_{2} e + (1-e) log_{2} (1-e)  ]\}
\end{equation}

Here transmission repetition rate is represented as $\nu$. The dark count probability, $p_{dark}$, is written as 

\begin{equation}
 p_{dark} =2d  
\end{equation}

The error rate expression is defined in Eq. (6), and Table II shows the values of $f(e)$. 
Under the condition, $p_{dark} \ll p_{signal} \ll 1 $, we obtain the values from Equation 16, $R_{dpsqkd} \approx \nu (1 - \frac{\mu}{N})p_{signal}$, in the absence of quantum memory, or $R_{dpsqkd} \approx \nu (1- 2\mu)p_{signal}$, in the presence of quantum memory. These are satisfying with \cite{gisin2004towards, lo2005decoy}.

\begin{figure}[!t]
\centering
\includegraphics[width=6.000in]{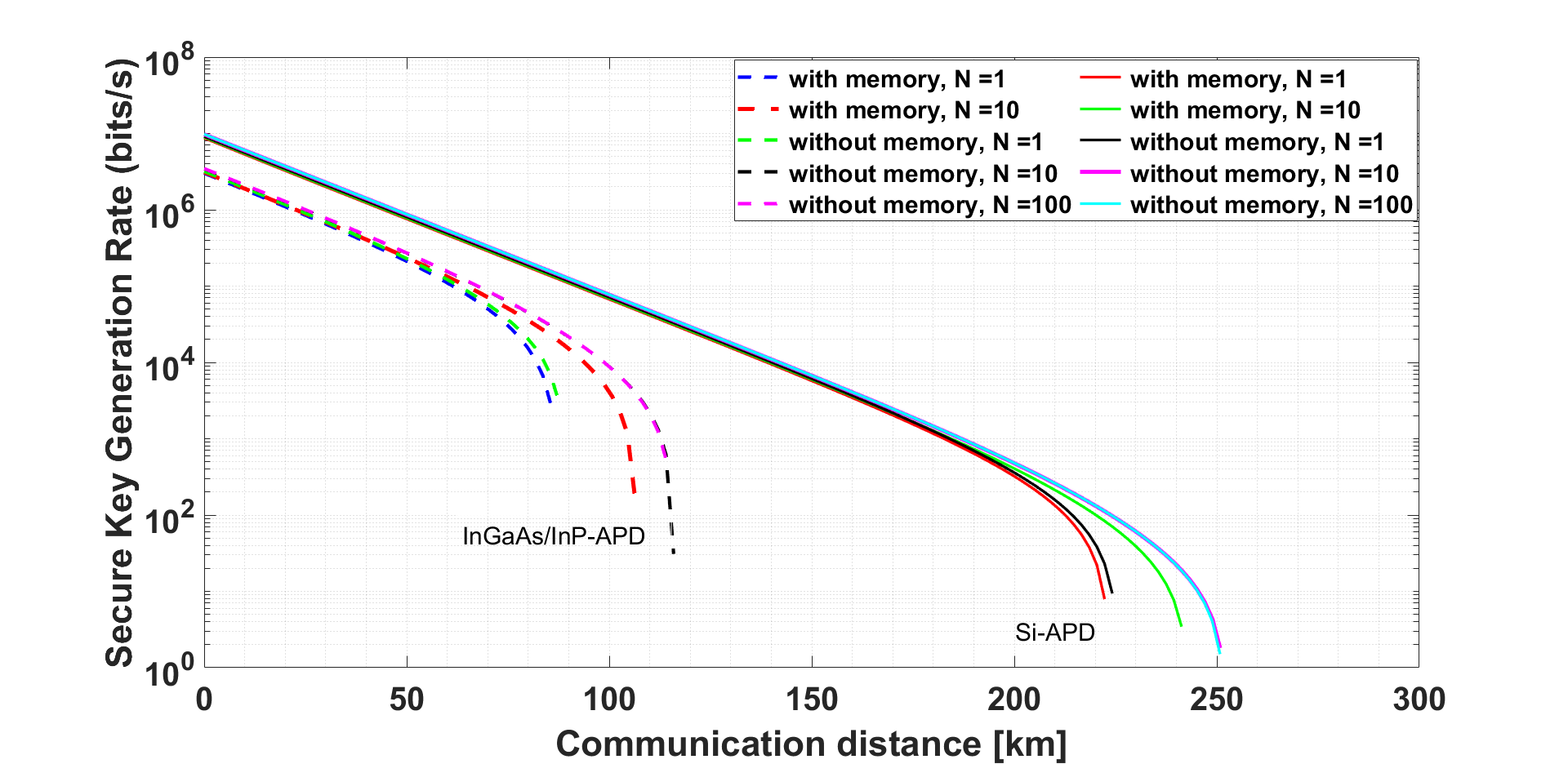}
\caption{\label{Rdpsqkd_24JUNE2022.png} Secure key generation under the considered attacks, $b = 0.01$; $\mu = 0.05$; $f=1.16$; $\nu=1*10^{9}$; $\eta_{1}=0.155$; $\eta_{2}=0.35$; $\alpha=0.21$; $L_{r1}=3.0$; $L_{r2}=2.1$; $d_{1}=9.2*10^{-6}$; $d_{2}=3.5*10^{-8}$; $N_{1}=1$; $N_{2}=10$; $N_{3}=100$; $t_{d1}=200*10^{-9}$; $t_{d2}=45*10^{-9}$. Here subscript 1 and 2 denote InGaAs-APD and Si-APD, respectively.}
\end{figure}

\begin{figure}[!t]
\centering
\includegraphics[width=7.000in]{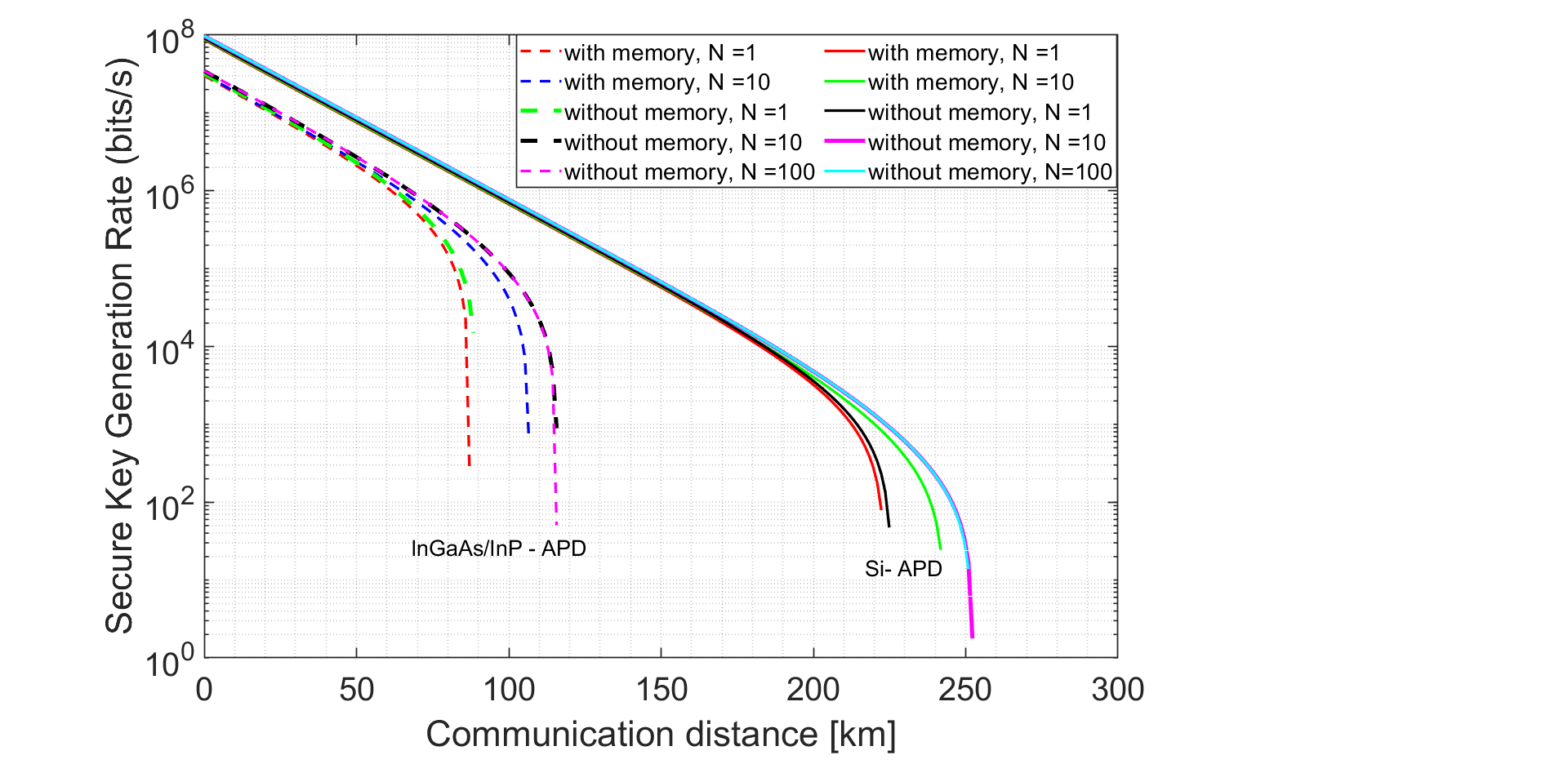}
\caption{\label{Rdpsqkd_24JUNE2022.png} Secure key generation under the considered attacks, $b = 0.01$; $\mu = 0.05$; $f=1.16$; $\nu=10*10^{9}$; $\eta_{1}=0.155$; $\eta_{2}=0.35$; $\alpha=0.21$; $L_{r1}=3.0$; $L_{r2}=2.1$;  $d_{1}=9.2*10^{-6}$; $d_{2}=3.5*10^{-8}$; $N_{1}=1$; $N_{2}=10$; $N_{3}=100$; $t_{d1}=200*10^{-9}$; $t_{d2}=45*10^{-9}$. Here subscript 1 and 2 denote InGaAs-APD and Si-APD, respectively.}
\end{figure}

\section{RESULTS  AND DISCUSSION}

\begin{figure}[!t]
\centering
\includegraphics[width=6.000in]{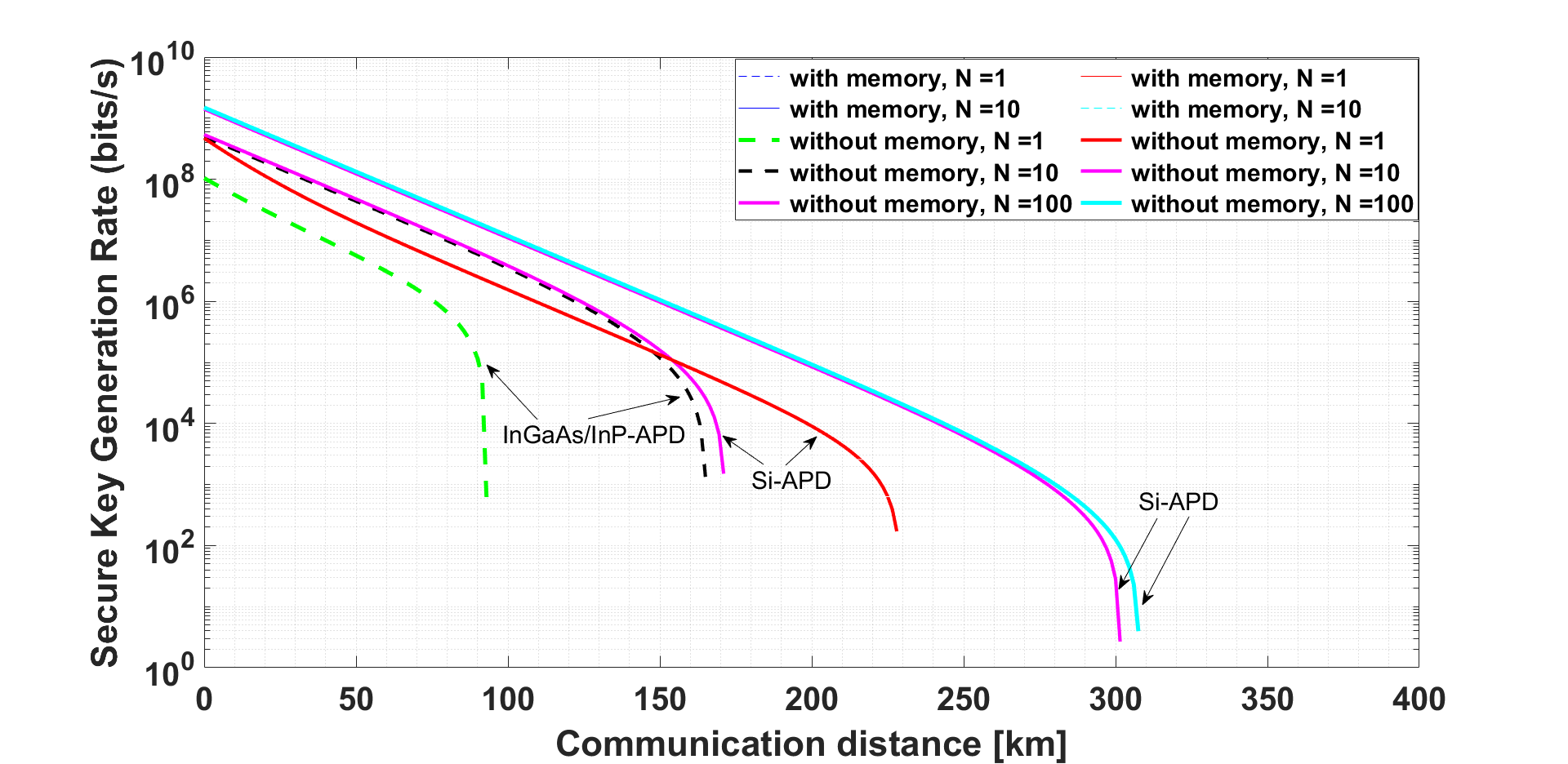}
\caption{\label{Rdpsqkd_24JUNE2022.png} Secure key generation under the considered attacks, $b = 0.01$; $\mu = 0.77$; $f=1.16$; $\nu=10*10^{9}$; $\eta_{1}=0.155$; $\eta_{2}=0.35$; $\alpha=0.21$; $L_{r1}=3.0$; $L_{r2}=2.1$;  $d_{1}=9.2*10^{-6}$; $d_{2}=3.5*10^{-8}$; $N_{1}=1$; $N_{2}=10$; $N_{3}=100$; $t_{d1}=200*10^{-9}$; $t_{d2}=45*10^{-9}$. Here subscript 1 and 2 denote InGaAs-APD and Si-APD, respectively.}
\end{figure}

With the described parameters and two types of detectors, we investigated the performance of Differential Phase shift QKD using frequency up-conversion. The values of the parameters under investigation are $\alpha = 0.2 dB/ km $ at 1550 nm, baseline system error rate, $b = 0.01$, extra loss at receiver end is $L_{r}$ = 1 dB. Here the secure communication rate in DPS-QKD  is optimized with respect to the value of the mean photon number, $\mu$. Too low value of $\mu$ generates high dark counts, and too high values of $\mu$ is responsible for photon-number splitting attack. 

\begin{figure}[!t]
\centering
\includegraphics[width=6.000in]{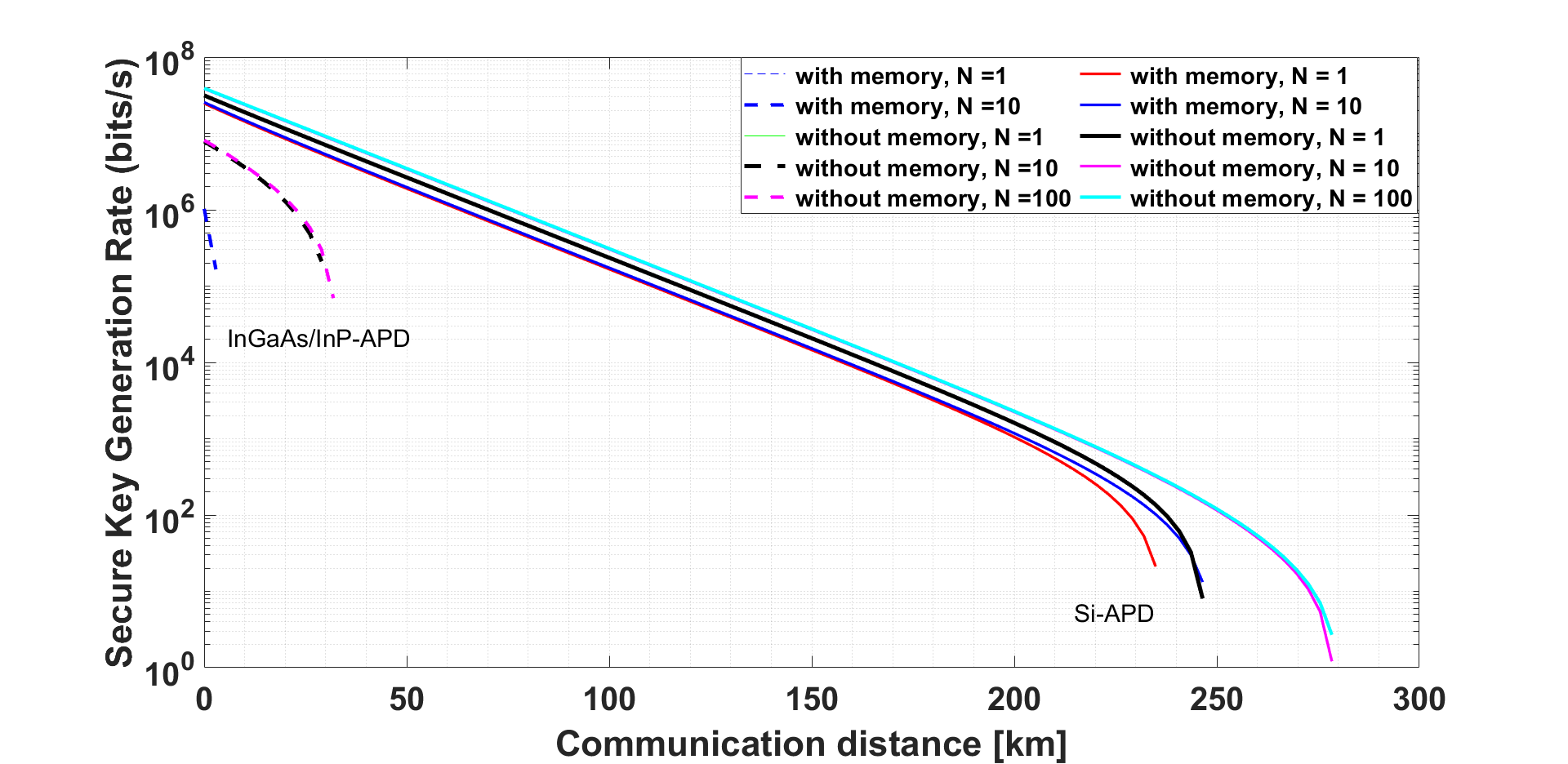}
\caption{\label{Rdpsqkd_24JUNE2022.png} Secure key generation under the considered attacks, $b = 0.01$; $\mu = 0.2$; $f=1.16$; $\nu=1*10^{9}$; $\eta_{1}=0.155$; $\eta_{2}=0.35$; $\alpha=0.21$; $L_{r1}=3.0$; $L_{r2}=2.1$;  $d_{1}=2.0*10^{-3}$; $d_{2}=3.5*10^{-8}$; $N_{1}=1$; $N_{2}=10$; $N_{3}=100$; $t_{d1}=200*10^{-9}$; $t_{d2}=45*10^{-9}$. Here subscript 1 and 2 denote InGaAs-APD and Si-APD, respectively.}
\end{figure}

It is clear from the simulated results obtained by Equation (16), and as shown in all the Figures (from Fig. 3 to Fig. 12) that the performance of the considered DPS-QKD protocol with two types of single photon detectors is greatly affected by detector quantum efficiency $\eta$, transmission repetition rate, $\nu$, dark counts,  $d$, and after pulse probability. Due to non-gated mode operation of Si-APD with considerable  timing jitter values, we achieve better results at 1 GHz and 10 GHz. The performance limiting factor is dead time, $t_{d}$,  in such cases. Here, in Poisson process, the two events occurring probability value in case of larger time $t_{d}$ is given as $e^{-\delta \nu p_{click}t_{d}}$, here the value of $\delta$ depends on the number of used detectors. Here, the value of $t_{d}$ = 45 ns, is taken into consideration. Using proper curve fitting method \cite{inoue2003differential}, we can tune the communication distance achieved with the pump power,  $p$. Hence, the obtained optimum results as shown  from Fig. 3 to Fig. 12 with the  consideration of all such situations and the values of different parameters  are mentioned in the caption of each Figures from Fig. 3 to Fig. 12 reflect the optimum results of secure key rates ranging from $10^{7}$ bits/sec to $10^{9}$ bits/sec, in the range of 200 km to 310 km communication distance.

Here, we observe that DPS-QKD performs efficiently under the PNS (photon-number splitting) attack, as described in the section of security analysis. In the simulated results we observe that, in the case when Eve has quantum memory, from Eq. (14), PNS attack does not affect the performance, the reason is that this attack is independent of delay term, $N$. On the opposite side, when Eve has no quantum memory with an infinitely long coherence time, the performance (secure key rate and communication distance) is severely affected, as shown in the simulated Figures. As shown in the results, when we tune the considered parameters, and delay parameter, $N$ ($N$ = 1, 10, 100),  we achieve secure key rates ranging from $10^{7}$ bits/sec to $10^{9}$ bits/sec, in the range of 200 km to 310 km communication distance. These optimum values of performance parameters of DPS-QKD prove its practical feasibility under all the considered conditions. Under two types of detectors used, Si-APD outperforms under frequency up-conversion in terms of considered performance parameters.

\section{Conclusion}
The simulated protocol achieves improved results in terms of secure key generation rate, which indicates that the two APDs under consideration performs much better by deploying frequency conversion technique in a Periodically poled lithium niobate (PPLN) waveguide. Moreover, it is clearly shown in the results that Si-APD with frequency up conversion provides enhanced communication distance and secure key generation rates as compared to InGaAs-APD which are also practically feasible in real field applications where single photon detection is required at telecommunication wavelength within the acceptable quantum bit error rates.
In the current research work, two types of single photon detectors at telecom wavelength are analyzed, one type of detector is InGaAs/ InP  APD and another one is based on frequency up-conversion method in PPLN waveguide and further detected by silicon APD. In addition to this, security analysis under certain attacks and communication rate equations are derived and analyzed for DPS-QKD protocol. Our simulation results clearly indicate that Si-APD with frequency up conversion outperforms InGaAs/ InP  APD under the considered parameter values and the hybrid attacks discussed.The improved simulation results  for sifted key rate, secure key rate and quantum bit error rates are clearly highlighted which proves that Si-APD with frequency up-conversion outperforms InGaAs/ InP  APD. Further, Eve’s capabilities were analyzed with and without memory, which affects the performance of the overall quantum communication system. Under such conditions, from the simulated results it is clearly observed that at high frequencies i.e. 1 GHz and 10 GHz it is possible to reach more than 300 km with considerably high secure key rates. To overcome the problems generated from birefringence and chromatic dispersion in fiber cables, it is required to deploy dispersion compensation  methods \cite{fasel2004quantum}  and  phase-encoding protocols \cite{honjo2004differential}.  These efforts will further improve the discussed performance parameters in realistic quantum communication scenarios.

\section{Acknowledgment}
Author acknowledges Indian Institute of Science, Bangalore for providing the support by the project Centre for Excellence in Quantum Technology (No. 4(7)/2020-ITEA), funded by the Ministry of Electronics and Information Technology, Government of India.

\textbf{Authors’ contributions} V.S. has directly participated in the planning, execution, and analysis of this study. V.S. drafted the manuscript. V.S. has read and approved the final version
of the manuscript. \newline
\newline
\textbf{Funding} V.S. would like to acknowledge Indian Institute of Science, Bangalore for providing the support by the project Centre for Excellence in Quantum Technology (No. 4(7)/2020-ITEA), funded by the Ministry of Electronics and Information Technology, Government of India. \newline
\newline
\textbf{Availability of data and materials} Not applicable. \newline
\newline
\textbf{Declarations} \newline
\newline
\textbf{Conflict of interest} There is no conflict of interest regarding the publication of this manuscript. \newline
\newline
\textbf{Consent for publication} Author is accepting to submit and publish the submitted work. \newline
\newline
\textbf{Ethical Approval} Not Applicable - The manuscript does not contain any human or animal studies. \newline
\newline
\textbf{Competing interests} The author declares that he has no competing interests.


\begin{thebibliography}{1}

\bibitem{bennett1992experimental}
Bennett, Charles H and Bessette, Fran{\c{c}}ois and Brassard, Gilles and Salvail, Louis and Smolin, John,
\newblock Experimental quantum cryptography
\newblock {\em J. Cryptology}, {\bf 5}, pages 3--28, (1992).

\bibitem{bennett1984proceedings}
Bennett, Charles H and Brassard, Gilles,
\newblock Proceedings of the ieee international conference on computers, systems and signal processing,
\newblock {\em IEEE New York}, {\bf 5}, pages 3--28, (1984).


\bibitem{honjo2004differential}
Honjo, T and Inoue, K and Takahashi, H,
\newblock Differential-phase-shift quanum key distribution experiment with a planar light-wave circuit Mach-Zehnder interferometer,
\newblock {\em Opt. Lett.}, {\bf 29}, pages 2797--2799, (2004). 



\bibitem{bennett1992quantum}
Bennett, Charles H and Brassard, Gilles and Mermin, N David,
\newblock  Quantum cryptography without Bell’s theorem,
\newblock {\em Physical review letters, APS},  {\bf 68 (5)}, pages 557, (1992). 
 


\bibitem{lutkenhaus2000security}
L{\"u}tkenhaus, Norbert,
\newblock Security against individual attacks for realistic quantum key distribution,
\newblock {\em Physical Review A, APS}, {\bf 61 (5)}, pages 052304, (2000).
  


\bibitem{waks2002security}
Waks, Edo and Zeevi, Assaf and Yamamoto, Yoshihisa,
\newblock  Security of quantum key distribution with entangled photons against individual attacks,
\newblock {\em Physical Review A, APS},  {\bf 65 (5)}, pages 052310, (2002).
  

\bibitem{inoue2002differential}
Inoue, Kyo and Waks, Edo and Yamamoto, Yoshihisa,
\newblock Differential phase shift quantum key distribution, 
\newblock {\em  Physical review letters, APS}, {\bf 89 (3)}, pages 037902, (2002).
  
 
\bibitem{inoue2003differential}
Inoue, K and Waks, E and Yamamoto, Y,
Differential-phase-shift quantum key distribution using coherent light,
\newblock {\em  Physical Review A, APS}, {\bf 68 (2)}, pages 022317, (2003).

\bibitem{yoshizawa200410}
Yoshizawa, Akio and Kaji, Ryosaku and Tsuchida, Hidemi,
\newblock  10.5 km fiber-optic quantum key distribution at 1550 nm with a key rate of 45 kHz,
\newblock {\em Japanese journal of applied physics, IOP Publishing}, {\bf 43 (6A)}, pages L735, (2004).
  

\bibitem{bethune2004high}
Bethune, Donald S and Risk, William P and Pabst, Gary W,
\newblock A high-performance integrated single-photon detector for telecom wavelengths,
\newblock {\em  Journal of modern optics, Taylor \& Francis}, {\bf 51 (9-10)}, pages 1359--1368, (2004).


\bibitem{stucki2001photon}
Stucki, Damien and Ribordy, Gr{\'e}goire and Stefanov, Andr{\'e} and Zbinden, Hugo and Rarity, John G and Wall, Tom,
\newblock Photon counting for quantum key distribution with Peltier cooled InGaAs/InP APDs,
\newblock {\em Journal of modern optics, Taylor \& Francis}, {\bf 48 (13)}, pages 1967--1981, (2001).


\bibitem{langrock2005highly}
Langrock, Carsten and Diamanti, Eleni and Roussev, Rostislav V and Yamamoto, Yoshihisa and Fejer, Martin M and Takesue, Hiroki, 
\newblock Highly efficient single-photon detection at communication wavelengths by use of upconversion in reverse-proton-exchanged periodically poled LiNbO waveguides,
\newblock {\em Optics letters, Optica Publishing Group}, {\bf 30 (13)}, pages 1725--1727, (2005).
  

\bibitem{bourennane2001single}
Bourennane, Mohamed and Karlsson, Anders and Ciscar, Juan Pena and Math{\'e}s, Markus,
\newblock Single-photon counters in the telecom wavelength region of 1550 nm for quantum information processing,
\newblock {\em Journal of modern optics, Taylor \& Francis}, {\bf 48 (13)}, pages 1983--1995, (2001).
 

\bibitem{gobby2004unconditionally}
Gobby, C and Yuan, ZL and Shields, AJ,
\newblock  Unconditionally secure quantum key distribution over 50km of standard telecom fibre,
\newblock {\em  arXiv preprint quant-ph/0412173}, (2004).
 

\bibitem{roussev2004periodically}
Roussev, Rostislav V and Langrock, Carsten and Kurz, Jonathan R and Fejer, Martin M,
\newblock  Periodically poled lithium niobate waveguide sum-frequency generator for efficient single-photon detection at communication wavelengths,
\newblock {\em Optics letters, Optica Publishing Group}, {\bf 29 (13)}, pages 1518--1520, (2004).
  

\bibitem{sharma2020quantum}
Sharma, Vishal and Banerjee, Subhashish,
\newblock Quantum communication using code division multiple access network,
\newblock {\em Optical and Quantum Electronics}, {\bf 52}(8), pages 1--22,  (2020).

\bibitem{gobby2004quantum}
Gobby, C and Yuan, aZL and Shields, AJ,
\newblock  Quantum key distribution over 122 km of standard telecom fiber
\newblock {\em Appl. Phys. Lett.}, {\bf 84}, pages 3762-–3764, (2004).

\bibitem{raj2019quantum}
Raj, Arockia Bazil and Sharma, Vishal and Banerjee, Subhashish,
\newblock  Quantum-based satellite free space optical communication and microwave photonics,
\newblock {\em Principles and applications of free space optical communications, IET Telecommunications Series}, {\bf 78}, (2019).

\bibitem{sharma2015controlled}
Sharma, Vishal and Shukla, Chitra and Banerjee, Subhashish and Pathak, Anirban,
\newblock Controlled bidirectional remote state preparation in noisy environment: a generalized view,
\newblock {\em Quantum Information Processing, Springer}, {\bf 14}, pages 3441--3464, (2015).
  
\bibitem{sharma2016comparative}
 Sharma, Vishal and Thapliyal, Kishore and Pathak, Anirban and Banerjee, Subhashish,
 \newblock  A comparative study of protocols for secure quantum communication under noisy environment: single-qubit-based protocols versus entangled-state-based protocols,
 \newblock {\em Quantum Information Processing, Springer},  {\bf 15}, pages 4681--4710, (2016).
  
\bibitem{sharma2018decoherence}
Sharma, Vishal and Shrikant, U and Srikanth, R and Banerjee, Subhashish,
\newblock Decoherence can help quantum cryptographic security,
\newblock {\em Quantum Information Processing, Springer}, {\bf 17}, pages 1--16, (2018).
  
\bibitem{sharma2018quantum}
Sharma, Vishal,
\newblock Quantum Communication Under Noisy Environment: From Theory to Applications,
\newblock {\em Indian Institute of Technology Jodhpur}, (2018).








\bibitem{sharma2018analysis}
Sharma, Vishal and Banerjee, Subhashish,
\newblock Analysis of quantum key distribution based satellite communication
\newblock {\em  In 2018 9th International Conference on Computing, Communication and Networking Technologies (ICCCNT)}, pp. 1-5. IEEE, 2018. 


\bibitem{inoue2003differential}
Inoue, K and Waks, E and Yamamoto, Y,
\newblock Differential-phase-shift quantum key distribution using coherent light
\newblock {\em Physical Review A}, {\bf 68}(2), pages 022317, (2003).

\bibitem{sharma2019analysis}
Sharma, Vishal and Banerjee, Subhashish,
\newblock Analysis of atmospheric effects on satellite-based quantum communication: a comparative study
\newblock {\em Quantum Information Processing}, {\bf 18}(3), pp. 1--24, 2019.

\bibitem{pelc2012cascaded}
Pelc, JS and Zhang, Q and Phillips, CR and Yu, L and Yamamoto, Y and Fejer, MM
\newblock Cascaded frequency upconversion for high-speed single-photon detection at 1550 nm
\newblock {\em Optics letters}, {\bf 37}(4), pages 476--478,  (2012).

\bibitem{sharma2014analysis}
Sharma, Vishal and Sharma, Richa,
\newblock  Analysis of spread spectrum in MATLAB
\newblock {\em International Journal of Scientific \& Engineering Research}, {\bf 5}(1), pp. 1899-1902, 2014.

\bibitem{IDQ}
https://www.idquantique.com/quantum-sensing/products/id100/.

\bibitem{sharma2016effect}
Sharma, Vishal,
\newblock  Effect of Noise on Practical Quantum Communication Systems
\newblock {\em Defence Science Journal}, {\bf 66}(2), (2016).


\bibitem{sharma2021quantum}
Sharma, Vishal and Gupta, Shantanu and Mehta, Gaurav and Lad, Bhupesh K
\newblock A quantum-based diagnostics approach for additive manufacturing machine
\newblock {\em IET Collaborative Intelligent Manufacturing}, {\bf 3}(2), pages 184--192,  (2021).

\bibitem{sharma2014feasibility}
Sharma, Vishal,
\newblock Feasibility of temperature sensors in railway coaches,
\newblock {\em Int. J. Sci. Eng. Res}, {\bf 5}(2), pages 881--884,  (2014).

\bibitem{sharma2015experimental}
Sharma, Vishal and Panchariya, PC,
\newblock Experimental use of electronic nose for odour detection,
\newblock {\em International Journal of Engineering Systems Modelling and Simulation, Inderscience Publishers (IEL)}, {\bf 7}(4), pages 238--243,  (2015).



\bibitem{inoue2005robustness}
Inoue, Kyo and Honjo, Toshimori
\newblock Robustness of differential-phase-shift quantum key distribution against photon-number-splitting attack
\newblock {\em Physical Review A}, {\bf 71}(4), pages 042305,  (2005).



\bibitem{honjo2007long}
Honjo, T and Takesue, H and Kamada, H and Nishida, Y and Tadanaga, O and Asobe, M and Inoue, K
\newblock Long-distance distribution of time-bin entangled photon pairs over 100 km using frequency up-conversion detectors
\newblock {\em Optics express}, {\bf 15}(21), pages 13957--13964,  (2007).


\bibitem{gisin2002quantum}
Gisin, Nicolas and Ribordy, Gr{\'e}goire and Tittel, Wolfgang and Zbinden, Hugo,
\newblock Quantum cryptography
\newblock {\em Reviews of modern physics}, {\bf 74}(1), pages 145--195, (2002). 


\bibitem{brassard1994advances}
Brassard, Gilles and Salvail, L,
\newblock  Advances in cryptology eurocrypt'93,
\newblock  {\em Lecture Notes in Computer Science}, {\bf 765},  pages 410--423,  (1994).


\bibitem{aureatechnology}
\newblock  https://www.aureatechnology.com/images/produits



\bibitem{gisin2004towards}
 Gisin, Nicolas and Ribordy, Gr{\'e}goire and Zbinden, Hugo and Stucki, Damien and Brunner, Nicolas and Scarani, Valerio,
 \newblock  Towards practical and fast quantum cryptography,
 \newblock {\em arXiv preprint quant-ph/0411022}, 2004.
 
\bibitem{lo2005decoy}
 Lo, Hoi-Kwong and Ma, Xiongfeng and Chen, Kai,
 \newblock Decoy state quantum key distribution,
 \newblock {\em Physical review letters}, {\bf 94(23)}, pages 230504, (2005).
  
\bibitem{fasel2004high}
Fasel, Sylvain and Alibart, Olivier and Tanzilli, Sebastien and Baldi, Pascal and Beveratos, Alexios and Gisin, Nicolas and Zbinden, Hugo,
\newblock High-quality asynchronous heralded single-photon source at telecom wavelength,
\newblock {\em New Journal of Physics, IOP Publishing}, {\bf 6(1)}, pages 163, (2004).

\bibitem{yoshizawa200410}
Yoshizawa, Akio and Kaji, Ryosaku and Tsuchida, Hidemi,
\newblock 10.5 km fiber-optic quantum key distribution at 1550 nm with a key rate of 45 kHz,
\newblock{ \em  Japanese journal of applied physics, IOP Publishing}, {\bf 43(6A)}, pages L735, (2004).
  
\bibitem{ma2005practical}
Ma, Xiongfeng and Qi, Bing and Zhao, Yi and Lo, Hoi-Kwong,
\newblock Practical decoy state for quantum key distribution,
\newblock { \em Physical Review A, APS}, {\bf 72(1)}, pages 012326, (2005).
  
\bibitem{fasel2004quantum}
Fasel, Sylvain and Gisin, Nicolas and Ribordy, Gr{\'e}goire and Zbinden, Hugo,
\newblock  Quantum key distribution over 30 km of standard fiber using energy-time entangled photon pairs: a comparison of two chromatic dispersion reduction methods,
\newblock { \em The European Physical Journal D-Atomic, Molecular, Optical and Plasma Physics, Springer}, {\bf 30(1)}, pages 143--148, (2004).
 

\bibitem{honjo2004differential}
Honjo, T and Inoue, K and Takahashi, H,
Differential-phase-shift quantum key distribution experiment with a planar light-wave circuit Mach--Zehnder interferometer,
\newblock { \em Optics letters, Optica Publishing Group}, {\bf 29(23)}, pages 2797--2799, (2004).
  

\bibitem{albota2004efficient}
Albota, Marius A and Wong, Franco NC,
Efficient single-photon counting at 1.55 $\mu$m by means of frequency upconversion,
 

\bibitem{vandevender2004high}
Vandevender, Aaron P and Kwiat, Paul G,
High efficiency single photon detection via frequency up-conversion,
\newblock { \em Journal of Modern Optics, Taylor \& Francis}, {\bf 51(9-10)}, pages 1433--1445, (2004).

\bibitem{sharma2022analysis}
Sharma, Vishal and Bhardwaj, Asha,
Analysis of Differential Phase Shift Quantum Key Distribution using single-photon detectors,
\newblock { \em 2022 International Conference on Numerical Simulation of Optoelectronic Devices (NUSOD), IEEE}, pages {17--18}, (2022).



\bibitem{scarani2004quantum}
Scarani, Valerio and Acin, Antonio and Ribordy, Gr{\'e}goire and Gisin, Nicolas,
Quantum cryptography protocols robust against photon number splitting attacks for weak laser pulse implementations,
\newblock { \em Physical review letters, APS}, {\bf 92(5)}, pages 057901, (2004).


\bibitem{wang2005beating}
Wang, Xiang-Bin,
Beating the photon-number-splitting attack in practical quantum cryptography,
\newblock { \em Physical review letters, APS}, {\bf 94(23)}, pages 230503, (2005).

\end{thebibliography}
\end{document}